\documentclass[preprint,aps]{revtex4}
\usepackage{amsmath}
\usepackage{graphicx}
\usepackage{bm}
\usepackage{fancyhdr}
\usepackage{amsmath}
\usepackage{graphicx}
\usepackage{amsfonts}
\usepackage{amsbsy}
\usepackage{amssymb}
\usepackage[mathscr]{eucal}

\newcommand{\beq}{\begin{equation}}
\newcommand{\eeq}{\end{equation}}
\newcommand{\ba}{\begin{array}}
\newcommand{\ea}{\end{array}}
\newcommand{\bea}{\begin{eqnarray}}
\newcommand{\eea}{\end{eqnarray}}
\newcommand{\bseq}{\begin{subequations}}
\newcommand{\eseq}{\end{subequations}}

\begin{document}

\title{Improvements to Kramers Turnover Theory}
\author{Eli Pollak }
\affiliation{Chemical Physics Department, Weizmann Institute of Science, 76100 Rehovoth,
Israel}
\email{eli.pollak@weizmann.ac.il}
\author{Joachim Ankerhold}
\affiliation{Institut f\"{u}r Theoretische Physik, Universit\"{a}t Ulm, Ulm, Germany}
\email{joachim.ankerhold@uni-ulm.de}
\date{\today}

\begin{abstract}
The Kramers turnover problem, that is obtaining a uniform expression for the rate of escape of a particle over a barrier for any value of the external friction was solved in the eighties. Two formulations were given, one by Mel'nikov and Meshkov (MM) (J. Chem. Phys. {\bf 85}, 1018 (1986)), which was based on a perturbation expansion for the motion of the particle in the presence of friction. The other, by Pollak, Grabert and H{\"a}nggi (PGH) (J. Chem. Phys. {\bf 91}, 4073 (1989)), valid also for memory friction, was based on a perturbation expansion for the motion along the collective unstable normal mode of the particle. Both theories did not take into account the temperature dependence of the average energy loss to the bath. Increasing the bath temperature will reduce the average energy loss. In this paper, we analyse this effect, using a novel perturbation theory. We find that within the MM approach, the thermal energy gained from the bath diverges, the average energy gain becomes infinite, implying an essential failure of the theory. Within the PGH approach increasing the bath temperature reduces the average energy loss but only by a finite small amount, of the order of the inverse of the reduced barrier height. This then does not seriously affect the theory. Analysis and application for a cubic potential and Ohmic friction are presented.
\end{abstract}

\maketitle






\section{\protect\bigskip Introduction}

One of the outstanding problems of the second half of the twentieth century in rate theory was the solution for the rate of escape of a particle trapped in a potential well over an adjacent barrier whose motion is subject to a Gaussian random force and associated frictional force. In 1940, Kramers \cite{kramers40} solved this problem in two limits. In the underdamped limit, when the friction is weak he found that the rate increases linearly with increasing friction. In the opposite strong damping limit, the rate decreases inversely with the friction strength. Kramers however did not manage to derive an expression for the rate which is valid for the whole range of friction. This problem is known as the Kramers turnover problem.

An important milestone in solution of this problem was given by Mel'nikov and Meshkov (MM) in 1986 \cite{melnikov86,melnikov91}. They introduced into the problem the concept of a conditional probability for the particle initiated at the barrier with energy $E$ to return to it with energy $E'$. Using perturbation theory with the friction strength as the small parameter they derived an explicit Gaussian expression for the kernel, in which the central quantity was the average energy lost by the particle to the bath as it traversed from the barrier to the well and back. This energy loss was temperature independent. Given the kernel, they wrote down a master equation for the energy flow and upon solving it were able to derive an expression for the depopulation factor rate which went smoothly from the weak damping limit where it was linear in the friction strength to unity for strong damping. They then wrote the full rate expression as a product of three terms, the "standard" transition state theory expression for the rate, the depopulation factor, and Kramers' spatial diffusion factor which went from unity at weak damping to being inversely proportional to the damping when the friction was strong.

Kramers' original theory dealt with Ohmic friction, that is the white noise case. Grote and Hynes noted in the early eighties that in Chemistry one should expect the environment to have memory, since the typical time scales of a molecular surrounding are similar to that of the reacting system. They then proceeded to solve the Kramers rate problem in the presence of memory friction, in the spatial diffusion limited regime \cite{grote80}, that is in the regime of moderate to strong friction. Carmeli and Nitzan \cite{carmeli82} proceeded to do the same for the underdamped regime. However the turnover problem in the presence of memory friction remained unsolved, even after the seminal result of Mel'nikov and Meshkov.

The solution for the turnover problem, without the need for the ansatz that the rate is a product of three terms, was derived in a few steps. First, Pollak \cite{pollak86} realised in 1986 that Kramers' problem in the spatial diffusion limit may be derived using variational transition state theory, based on a Hamiltonian description of the dynamics, in which the system is bilinearly coupled to a harmonic bath. In the vicinity of the barrier, the Hamiltonian is a quadratic form and so can be diagonalized. The normal modes then include an unstable collective mode and stable bath modes. The unstable mode is a linear combination of the system coordinate and the original (coupled) bath modes. Then Grabert \cite{grabert88}, using the discretized oscillator model derived a generalization of Mel'nikov and Meshkov's turnover theory by considering the dynamics of the unstable normal mode, instead of the system coordinate. The continuum limit version of this theory was then derived in Ref.  \cite{PGH} and is known as PGH theory. Compared to Mel'nikov and Meshkov's theory, PGH theory has two advantages: a) it is derived, there is no ansatz in the derivation; b) The result is also valid for memory friction. Practically though, the expression for the energy loss of the particle appearing in Mel'nikov and Meshkov's theory is much simpler to implement.

Both theories ignored the fact that the average energy loss should depend on the bath temperature. As the bath temperature increases, the fluctuations increase and the bath not only dissipates energy from the system due to the frictional force, it also pumps energy back into the system. As long as the energy of the system is much higher than the thermal energy ($k_BT$) one expects that the particle will on the average lose energy to the bath, however as the temperature is increased the average energy lost will be reduced. This is the topic of this present paper. We will use a perturbation expansion to derive a term for the energy gained by the particle during one traversal, using the PGH approach, based on the motion of the unstable normal mode and the MM approach, based on the motion of the system coordinate. We will show that for the MM theory the average energy gain diverges, since the particle feels the fluctuations of the bath during all of its traversal. This is then an essential failure of the MM theory. For PGH theory though, to leading order in the perturbation theory the average energy loss is a sum of two parts. One is the temperature independent energy loss, due to the frictional force. The other is a fluctuation induced energy gain which grows linearly with the temperature of the bath. This energy gain is finite, since in the vicinity of the barrier the unstable mode motion decouples from the bath. Inclusion of the energy gain term in the depopulation factor is then shown to lead to a correction which is inversely proportional to the reduced barrier height, and so typically, can be neglected.

In Section II we consider the average energy loss and energy gain within PGH theory. Then in Section III we consider the average energy loss and gain within MM theory and show that for Ohmic friction, the average energy gain, at any finite temperature, diverges. In Section IV we analyse the effect of the energy gain on the depopulation factor within PGH theory. A specific analytic example for a cubic oscillator and Ohmic friction is considered in Section V. We end with a discussion of implications and further improvements of PGH theory.

\renewcommand{\theequation}{2.\arabic{equation}} \setcounter{section}{1} %
\setcounter{equation}{0}

\section{The energy loss in PGH theory}

\subsection{Preliminaries}

The classical dynamics of the generic system is that of a particle with
mass $M$ and coordinate $q$\ whose classical equation of motion is a
Generalized Langevin Equation (GLE) of the form:%
\begin{equation}
M\ddot{q}+\frac{dV\left( q\right) }{dq}+M\int_{0}^{t}dt^{\prime }\gamma
\left( t-t^{\prime }\right) \dot{q}\left( t^{\prime }\right) =F\left(
t\right) .  \label{2.1}
\end{equation}%
$F\left( t\right) $ is a Gaussian random force with zero mean and
correlation function
\begin{equation}
\left\langle F\left( t\right) F\left( t^{\prime }\right) \right\rangle
=Mk_{B}T\gamma \left( t-t^{\prime }\right) .  \label{2.2}
\end{equation}%
$\gamma \left( t\right) $ is the friction function, $k_{B}$ is Boltzmann's
constant and $T$ is the temperature. The potential is assumed to have a well
at $q_{a}$ with frequency $\omega _{a}$ and a barrier at $q=0$ which
separates the well from either another well or a continuum. The harmonic
(imaginary) frequency at the barrier top is denoted as $\omega ^{\ddag }$.
The potential is then written as
\begin{equation}
V\left( q\right) =-\frac{1}{2}M\omega ^{\ddag 2}q^{2}+V_{1}\left( q\right)
\label{2.3}
\end{equation}%
and $V_{1}\left( q\right) $ is termed the nonlinear part of the potential
function. Kramers' turnover theory provides an expression for the rate of
escape of the particle from the well to the adjacent well or to the adjacent
continuum as a function of the friction function, the temperature, the
barrier height $V^{\ddag }=V\left( 0\right) -V\left( q_{a}\right) $, the
mass of the particle and properties of the potential.

It is well known \cite{zwanzig73} that the GLE may be derived as the continuum limit of the
equations of motion of the particle bilinearly coupled to a bath of harmonic
oscillators, with mass weighted coordinates $x_{j}$ and momenta $p_{j}$:
\begin{equation}
H=\frac{p^{2}}{2M}+V\left( q\right) +\frac{1}{2}\sum_{j=1}^{N}\left[
p_{j}^{2}+\omega _{j}^{2}\left( x_{j}-\frac{\sqrt{M}c_{j}}{\omega _{j}^{2}}%
q\right) ^{2}\right]   \label{2.4}
\end{equation}%
where $p$ is the momentum of the particle. The friction function is
identified as%
\begin{equation}
\gamma \left( t\right) =\sum_{j=1}^{N}\frac{c_{j}^{2}}{\omega _{j}^{2}}\cos
\left( \omega _{j}t\right)   \label{2.5}
\end{equation}%
and the random force is a function of the initial conditions of the bath
oscillators:%
\begin{equation}
F\left( t\right) =\sqrt{M}\sum_{j=1}^{N}c_{j}\left[ \left( x_{j,0}-\frac{%
\sqrt{M}c_{j}q_{0}}{\omega _{j}^{2}}\right) \cos \left( \omega _{j}t\right) +%
\frac{p_{j,0}}{\omega _{j}}\sin \left( \omega _{j}t\right) \right] .
\label{2.6}
\end{equation}%
Here, the $0$ subscript denotes the initial time. One readily finds that
averaging the random force with the canonical distribution $\exp \left(
-\beta H\right) $ with ($\beta =1/\left( k_{B}T\right) $) implies that it is
Gaussian with zero mean and correlation function as given in Eq. \ref{2.2}.

Around the barrier top, the Hamiltonian has a quadratic form and so may be
diagonalized. The details of this normal mode transformation may be found
for example in Ref. \cite{liao01}. The Hamiltonian is written in terms of the system
(unstable) mass weighted normal mode $\rho $ and momentum $p_{\rho }$ and stable
bath normal mode coordinates $y_{j}$ and momenta $p_{y_{j}}$ as:%
\begin{equation}
H=\frac{p_{\rho }^{2}}{2}-\frac{1}{2}\lambda ^{\ddag 2}\rho ^{2}+V_{1}\left(
q\right) +\frac{1}{2}\sum_{j=1}^{N}\left[ p_{y_{j}}^{2}+\lambda
_{j}^{2}y_{j}^{2}\right]   \label{2.7}
\end{equation}%
with $\lambda _{j}$, the frequency of the j-th normal mode. The system
coordinate $q$ is expressed in terms of the normal modes as
\begin{equation}
\sqrt{M}q=u_{00}\rho +\sum_{j=1}^{N}u_{j0}y_{j}  \label{2.8}
\end{equation}%
where $u_{j0}$ is the projection of the system coordinate on the j-th normal
mode. The nonlinear part of the potential $V_{1}\left( q\right) $ couples
the motion of the unstable normal mode $\left( \rho \right) $ motion to that
of the stable normal modes ($y_{j}$). The unstable normal mode barrier
frequency $\lambda ^{\ddag }$ is expressed in terms of the system barrier
frequency and Laplace transform of the time dependent friction ($\hat{\gamma}%
\left( s\right) $) via the Kramers-Grote-Hynes relation \cite{kramers40,grote80}:%
\begin{equation}
\lambda ^{\ddag 2}+\hat{\gamma}\left( \lambda ^{\ddag }\right) \lambda
^{\ddag }=\omega ^{\ddag 2}  \label{2.9}
\end{equation}%
The matrix element $u_{00}$\ for the projection of the system coordinate on
the unstable mode is also expressed in terms of the Laplace transform of the
time dependent friction \cite{PGH}:%
\begin{equation}
u_{00}^{2}=\left[ 1+\frac{1}{2}\left( \frac{\hat{\gamma}\left( \lambda
^{\ddag }\right) }{\lambda ^{\ddag }}+\frac{\partial \hat{\gamma}\left(
s\right) }{\partial s}|_{s=\lambda ^{\ddag }}\right) \right] ^{-1}.
\label{2.10}
\end{equation}%
These last two relations imply that the unstable mode barrier frequency and
the projection amplitude $u_{00}$ are known in the continuum limit.

The assumption of weak coupling between the system and the bath is expressed
as \cite{PGH}:
\begin{equation}
u_{1}^{2}=1-u_{00}^{2}\ll 1  \label{2.11}
\end{equation}%
which equivalently implies that the normal mode transformation elements $%
u_{j0},j\neq 0$ are small.

The equation of motion for each of the stable normal modes is:%
\begin{equation}
\ddot{y}_{j}+\lambda _{j}^{2}y_{j}=-\frac{u_{j0}}{\sqrt{M}}V_{1}^{\prime
}\left( \frac{1}{\sqrt{M}}\left( u_{00}\rho
+\sum_{j=1}^{N}u_{j0}y_{j}\right) \right)   \label{2.12}
\end{equation}%
while the equation of motion for the unstable normal mode is:%
\begin{equation}
\ddot{\rho}-\lambda ^{\ddag 2}\rho =-\frac{u_{00}}{\sqrt{M}}V_{1}^{\prime
}\left( \frac{1}{\sqrt{M}}\left( u_{00}\rho
+\sum_{j=1}^{N}u_{j0}y_{j}\right) \right) .  \label{2.13}
\end{equation}

\subsection{\protect\bigskip Perturbation theory}

The motion for each of the stable bath modes may be represented in terms of
a power series in the small parameter $u_{j0}$. For our purposes it suffices
to consider the first two terms in such a series, so that for example the
coordinate of the j-th stable mode has two components $y_{j0,t}$ and $%
y_{j1,t}$ of order $u_{j0}^{0}$ and $u_{j0}^{1}$ respectively. Similarly the
system unstable mode motion may be expanded in a perturbation series such
that the first two terms for the coordinate of the unstable mode $\rho _{0,t}
$ and $\rho _{1,t}\,$ are of the order $u_{1}^{0}$ and $u_{1}^{1}$
respectively. The equations of motion for the components of each of the
stable modes is

\begin{eqnarray}
\ddot{y}_{j0,t}+\lambda _{j}^{2}y_{j0,t} &=&0  \label{2.14} \\
\ddot{y}_{j1,t}+\lambda _{j}^{2}y_{j1,t} &=&-\frac{u_{j0}}{\sqrt{M}}%
V_{1}^{\prime }\left( \frac{u_{00}\rho _{0,t}}{\sqrt{M}}\right)
\label{2.15}
\end{eqnarray}%
Similarly the equations of motion for the first two components of the
unstable normal mode are
\begin{equation}
\ddot{\rho}_{0,t}-\lambda ^{\ddag 2}\rho _{0,t}=-\frac{u_{00}}{\sqrt{M}}%
V_{1}^{\prime }\left( \frac{u_{00}\rho _{0,t}}{\sqrt{M}}\right)
\label{2.16a}
\end{equation}%
and
\begin{equation}
\ddot{\rho}_{1,t}-\lambda ^{\ddag 2}\rho _{1,t}=-\frac{u_{00}}{M}%
V_{1}^{\prime \prime }\left( \frac{u_{00}\rho _{0,t}}{\sqrt{M}}\right)
\left( u_{00}\rho _{1,t}+\sum_{j=1}^{N}u_{j0}y_{j0,t}\right) .  \label{2.17}
\end{equation}

A central quantity in Kramers' turnover theory is the energy gained by the
bath as the unstable mode motion moves from the barrier region to the well
and back. The stable mode bath energy is by definition%
\begin{equation}
E_{B}=\frac{1}{2}\sum_{j=1}^{N}\left( p_{y_{j}}^{2}+\lambda
_{j}^{2}y_{j}^{2}\right) .  \label{2.18}
\end{equation}%
The change in time of the bath energy including all terms up to $u_{j0}^{2}$
is readily seen to be:%
\begin{eqnarray}
\frac{dE_{B}}{dt} &=&\sum_{j=1}^{N}\left[ -\frac{u_{j0}\left( \dot{y}_{j0,t}+%
\dot{y}_{j1,t}\right) }{\sqrt{M}}V_{1}^{\prime }\left( \frac{u_{00}\rho
_{0,t}}{\sqrt{M}}\right) -\frac{u_{j0}}{M}\dot{y}_{j0,t}\left( u_{00}\rho
_{1,t}+\sum_{j=1}^{N}u_{j0}y_{j0,t}\right) V_{1}^{\prime \prime }\left(
\frac{u_{00}\rho _{0,t}}{\sqrt{M}}\right) \right]   \notag \\
&&+O\left( u_{j0}^{3}\right)   \label{2.19}
\end{eqnarray}%
so that to this order it suffices to consider only the zero-th and first
order solutions for both the stable modes and the unstable mode.

\bigskip The solution for the zero-th order equation of motion for the j-th
stable mode is that of a harmonic oscillator%
\begin{equation}
y_{j,0}\left( t\right) =y_{j,0}\left( t_{0}\right) \cos \left[ \lambda
_{j}\left( t-t_{0}\right) \right] +\frac{\dot{y}_{j,0}\left( t_{0}\right) }{%
\lambda _{j}}\sin \left[ \lambda _{j}\left( t-t_{0}\right) \right] .
\label{2.20}
\end{equation}%
It is useful to define a Gaussian random "noise function" $\Phi \left(
t\right) $ as:%
\begin{equation}
\Phi \left( t\right) =\sum_{j=1}^{N}u_{j0}y_{j0,t}.  \label{2.21}
\end{equation}%
When averaging over the uncoupled bath Hamiltonian using the thermal
distribution $\exp \left( -\beta E_{B}\left( t_{0}\right) \right) $ one
readily notes that the mean of $\Phi \left( t\right) $\ vanishes and the
noise correlation function is
\begin{equation}
\left\langle \Phi \left( t\right) \Phi \left( t^{\prime }\right)
\right\rangle =\frac{1}{\beta }\sum_{j=1}^{N}\frac{u_{j0}^{2}}{\lambda
_{j}^{2}}\cos \left[ \lambda _{j}\left( t-t^{\prime }\right) \right] \equiv
\frac{1}{\beta }K\left( t-t^{\prime }\right)   \label{2.22}
\end{equation}%
and this defines the "normal mode friction kernel" $K\left( t\right) $.
Using properties of the normal mode transformation (see for example Eq. 2.17
of Ref. \cite{liao01}) one may readily
express the Laplace transform of the kernel as%
\begin{equation}
\hat{K}\left( s\right) =\left( \frac{su_{00}^{2}}{\lambda ^{\ddag 2}\left(
s^{2}-\lambda ^{\ddag 2}\right) }+\frac{s+\hat{\gamma}\left( s\right) }{%
\omega ^{\ddag 2}\left( \omega ^{\ddag 2}-s^{2}-\hat{\gamma}\left( s\right)
s\right) }\right)   \label{2.23}
\end{equation}%
showing that it is defined in terms of the Laplace transform of the friction
function and thus in the continuum limit.

The solution for the first order equation of the j-th stable mode is (taking
$t_{0}\rightarrow -\infty )$:%
\begin{equation}
y_{j1,t}=-\frac{u_{j0}}{\sqrt{M}}\int_{-\infty }^{t}dt^{\prime }\frac{\sin %
\left[ \lambda _{j}\left( t-t^{\prime }\right) \right] }{\lambda _{j}}%
V_{1}^{\prime }\left( \frac{u_{00}\rho _{0,t^{\prime }}}{\sqrt{M}}\right)
\label{2.24}
\end{equation}%
and it is independent of the noise $\Phi $.

The zero-th order equation of motion for the unstable mode is that of a
conservative system with a nonlinear potential. In some cases, the
trajectory for the motion from the barrier top to the well and back to the
barrier top is known analytically. In any case, the period of this orbit is
infinite, it starts in the infinite past at the barrier and returns to it in
the infinite future. Henceforth we will limit all solutions to these
conditions, so that for example the initial time is taken as $-\infty $. The
solution for the first order equation of motion for the unstable mode is
somewhat more complex, however readily solved by using the conservation of
the overall energy of the system and the bath, as shown in Appendix A. One
finds that it is linear in the noise:
\begin{equation}
\rho _{1,t}=-\dot{\rho}_{0,t}\int_{-\infty }^{t}dt^{\prime }\frac{1}{\dot{%
\rho}_{0,t^{\prime }}^{2}}\int_{-\infty }^{t^{\prime }}dt^{\prime \prime }%
\frac{\Phi \left( t^{\prime \prime }\right) }{\sqrt{M}}\frac{d}{dt^{\prime
\prime }}V_{1}^{\prime }\left( \frac{u_{00}\rho _{0,t^{\prime \prime }}}{%
\sqrt{M}}\right) .  \label{2.25}
\end{equation}%
For later purposes, we note that the thermal average of the product of the first order
solution with the time derivative of the noise is:%
\begin{equation}
\left\langle \rho _{1,t}\dot{\Phi}\left( t\right) \right\rangle =-\frac{\dot{%
\rho}_{0,t}}{\sqrt{M}\beta }\int_{-\infty }^{t}dt^{\prime }\frac{1}{\dot{\rho%
}_{0,t^{\prime }}^{2}}\int_{-\infty }^{t^{\prime }}dt^{\prime \prime }\frac{%
dK\left( t-t^{\prime \prime }\right) }{dt}\frac{d}{dt^{\prime \prime }}%
V_{1}^{\prime }\left( \frac{u_{00}\rho _{0}\left( t^{\prime \prime }\right)
}{\sqrt{M}}\right) .  \label{2.26}
\end{equation}

\subsection{The thermally averaged energy loss and its variance}

\bigskip The thermally averaged energy loss is obtained by averaging the
energy change over the thermally distributed initial conditions of the bath $%
\exp \left( -\beta E_{B}\left( -\infty \right) \right) $. This implies that
the thermal averages of the stable mode coordinates and velocities vanish.
The coordinate correlation functions are:%
\begin{equation}
\left\langle y_{j0,t}y_{k0,t^{\prime }}\right\rangle =\frac{\cos \left[
\lambda _{j}\left( t-t^{\prime }\right) \right] }{\beta \lambda _{j}^{2}}%
\delta _{jk}  \label{2.27}
\end{equation}%
where $\delta _{jk}$ is the Kronecker "delta" function and from this
relation one obtains by differentiation with respect to the time the
velocity correlation functions.  This implies for example, that $%
\left\langle \dot{\Phi}\left( t\right) \Phi \left( t\right) \right\rangle =0$%
.

The average energy loss of the system, defined as the gain in the energy of
the bath as the unstable mode motion goes through one cycle, starting at the
barrier and returning to it, is obtained by integration of Eq. \ref{2.19}
over the unperturbed barrier top trajectory for the unstable mode, that is the
time goes from $-\infty $ to $\infty $. It is then readily found to be
\begin{equation}
\left\langle \Delta E\right\rangle =\int_{-\infty }^{\infty }dt\left\langle
\frac{dE_{B}}{dt}\right\rangle \equiv DE_{1}-DE_{2}.  \label{2.28}
\end{equation}%
Using the first order solution for the stable mode motion (Eq. \ref{2.24})
one finds that the systematic component of the energy loss $DE_{1}$ is:%
\begin{equation}
DE_{1}\equiv \frac{1}{2M}\int_{-\infty }^{\infty }dt\int_{-\infty }^{\infty
}dt^{\prime }V_{1}^{\prime }\left( \frac{u_{00}\rho _{0,t}}{\sqrt{M}}\right)
\frac{\partial ^{2}K\left( t-t^{\prime }\right) }{\partial t\partial
t^{\prime }}V_{1}^{\prime }\left( \frac{u_{00}\rho _{0,t^{\prime }}}{\sqrt{M}%
}\right)   \label{2.29}
\end{equation}%
and this is the term that also appears in PGH theory. The new result is that
to the same order in the perturbation strength, there is also a temperature
dependent contribution to the energy loss:
\begin{equation}
DE_{2}\equiv -\frac{u_{00}^{2}}{M^{2}\beta }\int_{-\infty }^{\infty
}dt\int_{-\infty }^{\infty }dt^{\prime \prime }V_{1}^{\prime \prime }\left(
\frac{u_{00}\rho _{0,t}}{\sqrt{M}}\right) \dot{\rho}_{0,t}\frac{dK\left(
t-t^{\prime \prime }\right) }{dt}V_{1}^{\prime \prime }\left( \frac{%
u_{00}\rho _{0,t^{\prime \prime }}}{\sqrt{M}}\right) \dot{\rho}_{0,t^{\prime
\prime }}\int_{t^{\prime \prime }}^{t}dt^{\prime }\frac{1}{\dot{\rho}%
_{0,t^{\prime }}^{2}}  \label{2.30}
\end{equation}%
which expresses the fact that when the bath is heated it may also transfer
energy to the system. It is this temperature dependent reheating of the
system which is missing in PGH\ theory.

The variance of the energy loss is by definition%
\begin{equation}
\sigma ^{2}=\left\langle \left( \Delta E-\left\langle \Delta E\right\rangle
\right) ^{2}\right\rangle .  \label{2.31}
\end{equation}%
To leading order we use only those terms which are up to
quadratic in $u_{j0}$. This implies that the term $\left\langle \Delta
E\right\rangle ^{2}$ which is fourth order in $u_{j0}$ may be ignored. The
leading order term in the energy loss is then identical to that
found in PGH theory, that is:
\begin{equation}
\sigma ^{2}=\frac{2}{\beta }DE_{1}.  \label{2.32}
\end{equation}
\bigskip \renewcommand{\theequation}{3.\arabic{equation}} %
\setcounter{section}{2} \setcounter{equation}{0}

\section{The energy loss for the system coordinate}

The turnover theory of Mel'nikov and Meshkov was based on a perturbation
expansion in which one considers directly the motion of the system
coordinate. The small parameter is taken to be the noise $F\left( t\right) $
and we note that the friction function is second order in the noise. The
system motion is separated into three parts%
\begin{equation}
q_{t}=q_{0,t}+q_{1,t}+q_{2,t}+O\left( F^{3}\right) .  \label{3.1}
\end{equation}%
which are of the order $F^{0},F^{1}$ and $F^{2}$ respectively. They obey the
equations of motion:%
\begin{eqnarray}
0 &=&M\ddot{q}_{0,t}+V^{\prime }\left( q_{0,t}\right) ,  \label{3.2} \\
F\left( t\right)  &=&M\ddot{q}_{1,t}+V^{\prime \prime }\left( q_{0,t}\right)
q_{1,t}  \label{3.3} \\
0 &=&M\ddot{q}_{2,t}+V^{\prime \prime }\left( q_{0,t}\right) q_{2,t}+\frac{1%
}{2}V^{\left( 3\right) }\left( q_{0,t}\right)
q_{1,t}^{2}+M\int_{-t_{0}}^{t}dt^{\prime }\gamma \left( t-t^{\prime }\right)
\dot{q}_{0,t^{\prime }}.  \label{3.4}
\end{eqnarray}%
The time derivative of the kinetic energy of the system, including all terms
up to order $F^{2}$ is then:%
\begin{eqnarray}
\frac{dE}{dt} &=&\dot{q}_{0,t}\left[ -V^{\prime }\left( q_{0,t}\right)
-V^{\prime \prime }\left( q_{0,t}\right) \left( q_{1,t}+q_{2,t}\right) -%
\frac{1}{2}V^{\left( 3\right) }\left( q_{0,t}\right)
q_{1,t}^{2}-M\int_{-t_{0}}^{t}dt^{\prime }\gamma \left( t-t^{\prime }\right)
\dot{q}_{0,t^{\prime }}+F\left( t\right) \right]   \notag \\
&&+\dot{q}_{1,t}\left[ -V^{\prime }\left( q_{0,t}\right) -V^{\prime \prime
}\left( q_{0,t}\right) q_{1,t}+F\left( t\right) \right] -\dot{q}%
_{t,2}V^{\prime }\left( q_{0,t}\right) +O\left( F^{3}\right)   \label{3.5}
\end{eqnarray}%
The energy loss is obtained by integrating using the zero-th order trajectory
that is initiated at the barrier and returns to the barrier. Using the
perturbative equations of motion given in Eqs. \ref{3.2}-\ref{3.4} one finds
that to order $F^2$ the energy loss depends only on the zero-th and first order
contributions:%
\begin{eqnarray}
-\Delta E_{q} &=&\int_{-\infty }^{\infty }dt\frac{dE}{dt}  \notag \\
&=&\int_{-\infty }^{\infty }dt\left[ -M\dot{q}_{0,t}\int_{-\infty
}^{t}dt^{\prime }\gamma \left( t-t^{\prime }\right) \dot{q}_{0,t^{\prime
}}+\left( \dot{q}_{0,t}+\dot{q}_{1,t}\right) F\left( t\right) \right]
+O\left( F^{3}\right) .  \label{3.6}
\end{eqnarray}%
Averaging over the random force then gives two contributions:%
\begin{equation}
-\left\langle \Delta E_{q}\right\rangle =-\int_{-\infty }^{\infty }dt\left[ M%
\dot{q}_{0,t}\int_{-\infty }^{t}dt^{\prime }\gamma \left( t-t^{\prime
}\right) \dot{q}_{0,t^{\prime }}+\left\langle \dot{q}_{1,t}F\left( t\right)
\right\rangle \right] +O\left( F^{3}\right) .  \label{3.7}
\end{equation}%
The first contribution on the right hand side, which is independent of
temperature, is the same as derived in the turnover theory of Mel'nikov and
Meshkov. The second contribution is though missing.

As in the PGH\ approach, one may derive an explicit expression also for the
second temperature dependent contribution to the energy loss. This is
readily carried out using the Hamiltonian representation for the Langevin equation,
as given in Eq. \ref{2.4}. For this purpose, the motion of each bath mode is
also treated with perturbation theory, where the small parameter is the
coupling constant $c_{j}$. The zero-th and first order contributions to the
motion of the j-th bath mode are then denoted as $x_{j0,t}$ and $x_{j1,t}$
respectively. The zero-th order motion is that of the uncoupled j-th
bath oscillator. The first order correction to \ the j-th bath oscillator
motion is readily found to be:%
\begin{equation}
x_{j1,t}=\sqrt{M}c_{j}\int_{-\infty }^{t}dt^{\prime }\frac{\sin \left[
\omega _{j}\left( t-t^{\prime }\right) \right] }{\omega _{j}}q_{0,t^{\prime
}}.  \label{3.8}
\end{equation}

Following the same derivation as in Appendix A, one notes that energy
conservation implies that to first order in the coupling coefficients $c_{j}$%
:
\begin{equation}
0=\frac{p_{q,0}p_{q,1}}{M}+V^{\prime }\left( q_{0}\right)
q_{1}+\sum_{j=1}^{N}\left[ p_{j0}p_{j1}+\omega _{j}^{2}x_{j0}x_{j1}\right]
-Fq_{0}.  \label{3.9}
\end{equation}%
This then gives a first order in time equation of motion for the first order
correction to the system coordinate which is readily solved:%
\begin{equation}
q_{1,t}=\dot{q}_{0,t}\int_{-\infty }^{t}dt^{\prime }\frac{1}{M\dot{q}%
_{0,t^{\prime }}^{2}}\int_{-\infty }^{t^{\prime }}dt^{\prime \prime }\dot{q}%
_{0,t^{\prime \prime }}F\left( t^{\prime \prime }\right)   \label{3.10}
\end{equation}%
We thus find that:%
\begin{eqnarray}
\left\langle \dot{q}_{1,t}F\left( t\right) \right\rangle  &=&\frac{M}{\beta }%
\frac{d\ln \dot{q}_{0}}{dt}\dot{q}_{0,t}\int_{-\infty }^{t}dt^{\prime }\frac{%
1}{M\dot{q}_{0,t^{\prime }}^{2}}\int_{-\infty }^{t^{\prime }}dt^{\prime
\prime }\dot{q}_{0,t^{\prime \prime }}\gamma \left( t-t^{\prime \prime
}\right)   \notag \\
&&+\frac{1}{\beta \dot{q}_{0,t}}\int_{-\infty }^{t}dt^{\prime }\dot{q}%
_{0,t^{\prime }}\gamma \left( t-t^{\prime }\right) .  \label{3.11}
\end{eqnarray}%
The energy gained by the system from the bath is thus:%
\begin{eqnarray}
\left\langle \Delta E_{q}\right\rangle _{2} &=&\int_{-\infty }^{\infty
}dt\left\langle \dot{q}_{t,1}F\left( t\right) \right\rangle   \notag \\
&=&\frac{M}{\beta }\int_{-\infty }^{\infty }dt\ddot{q}_{0,t}\int_{-\infty
}^{\infty }dt^{\prime \prime }\dot{q}_{0,t^{\prime \prime }}\gamma \left(
t-t^{\prime \prime }\right) \int_{t^{\prime \prime }}^{t}dt^{\prime }\frac{1%
}{M\dot{q}_{0,t^{\prime }}^{2}}  \notag \\
&&+\int_{-\infty }^{\infty }dt\frac{1}{\beta \dot{q}_{0,t}}\int_{-\infty
}^{t}dt^{\prime }\dot{q}_{0,t^{\prime }}\gamma \left( t-t^{\prime }\right) .
\label{3.12}
\end{eqnarray}%
Consider then the case of Ohmic friction, that is:%
\begin{equation}
\gamma \left( t-t^{\prime }\right) =2\gamma \delta \left( t-t^{\prime
}\right)   \label{3.13}
\end{equation}%
where $\delta \left( x\right) $ is the Dirac "delta" function. The first
term on the right hand side of the second equality in Eq. \ref{3.12}
vanishes but the second term diverges:%
\begin{equation}
\left\langle \Delta E_{q}\right\rangle _{2,Ohmic}=\frac{\gamma }{\beta }%
\int_{-\infty }^{\infty }dt\rightarrow \infty .  \label{3.14}
\end{equation}

This is then an essential failure of the Mel'nikov and Meshkov solution for
the turnover problem. Thermal fluctuations of the bath give an infinite
amount of energy to the system. The same divergence does not occur in
PGH theory where one considers the motion of the unstable normal mode. In
the vicinity of the barrier, the unstable mode is decoupled from the bath
and the bath can no longer affect the motion. In the Mel'nikov Meshkov
theory, even arbitrarily close to the barrier top, the system coordinate
remains coupled to the bath. Due to the infinite time duration of the motion, the
bath can then provide an infinite amount of energy and this causes the
divergence.

\renewcommand{\theequation}{4.\arabic{equation}} \setcounter{section}{3} %
\setcounter{equation}{0}

\section{PGH\ turnover theory}

\subsection{The energy loss conditional probability}

A central object in the turnover theory is the probability kernel $P\left(
E^{\prime }|E\right) $ that the system originates at the barrier with energy
$E$ and returns to it with energy $E^{\prime }$. From Eq. \ref{2.19} one
readily finds that the change in energy may be written as:%
\begin{eqnarray}
\Delta E &=&E-E^{\prime }=DE_{1}+\int_{-\infty }^{\infty }dtW_{1}\left[ \dot{%
\Phi}\left( t\right) \right]   \notag \\
&&+\int_{-\infty }^{\infty }dt\int_{-\infty }^{t}dt^{\prime }W_{2}\left[
\dot{\Phi}\left( t\right) ,\Phi \left( t^{\prime }\right) \right]
-\int_{-\infty }^{\infty }dtW_{3}\left[ \dot{\Phi}\left( t\right) ,\Phi
\left( t\right) \right]   \label{4.1}
\end{eqnarray}%
with
\begin{eqnarray}
W_{1}\left[ \dot{\Phi}\left( t\right) \right]  &=&-\frac{1}{\sqrt{M}}%
V_{1}^{\prime }\left( \frac{u_{00}\rho _{0}\left( t\right) }{\sqrt{M}}%
\right) \dot{\Phi}\left( t\right)   \label{4.2} \\
W_{2}\left[ \dot{\Phi}\left( t\right) ,\Phi \left( t^{\prime }\right) \right]
&=&\frac{1}{M}\dot{\Phi}\left( t\right) \frac{d}{dt}\left[ V_{1}^{^{\prime
}}\left( \frac{u_{00}\rho _{0}}{\sqrt{M}}\right) \right] \Phi \left(
t^{\prime }\right) \frac{d}{dt^{\prime }}V_{1}^{\prime }\left( \frac{%
u_{00}\rho _{0}\left( t^{\prime }\right) }{\sqrt{M}}\right) \int_{t^{\prime
}}^{t}dt^{\prime \prime }\frac{1}{\dot{\rho}_{0,t^{\prime \prime }}^{2}}
\notag \\ \label{4.3} \\
W_{3}\left[ \dot{\Phi}\left( t\right) ,\Phi \left( t\right) \right]  &=&%
\frac{\dot{\Phi}\left( t\right) \Phi \left( t\right) }{M}V_{1}^{\prime
\prime }\left( \frac{u_{00}\rho _{0}}{\sqrt{M}}\right)   \label{4.4}
\end{eqnarray}%
The thermally averaged probability $P\left( E^{\prime }|E\right) $ that the
particle ends with energy $E^{\prime }$ after being initiated with energy $E$
is then by definition%
\begin{eqnarray}
&&P\left( E^{\prime }|E\right) =  \notag \\
&&\left\langle \delta \left( E^{\prime }-E+DE_{1}+\int_{-\infty }^{\infty }dt
\left[ W_{1}\left[ \dot{\Phi}\left( t\right) \right] -W_{3}\left[ \dot{\Phi}%
\left( t\right) ,\Phi \left( t\right) \right] +\int_{-\infty }^{t}dt^{\prime
}W_{2}\left[ \dot{\Phi}\left( t\right) ,\Phi \left( t^{\prime }\right) %
\right] \right] \right) \right\rangle _{\beta }  \notag \\
&&  \label{4.5}
\end{eqnarray}%
where the average is over the distribution
\begin{equation}
P\left( \text{\b{y}},\text{\b{p}}_{y}\right) =\prod\limits_{j=1}^{N}\frac{%
\beta \lambda _{j}}{2\pi }\exp \left( -\frac{\beta }{2}\left( \dot{y}%
_{j0}^{2}+\lambda _{j}^{2}y_{j0}^{2}\right) \right) .  \label{4.6}
\end{equation}%
with $p_{y_{j}}=\dot{y}_{j}$ and we used the shortened notation $%
y_{j0}=y_{j,0}\left( t_{0}\right) $, etc..

If one considers only the first two terms in the energy loss (ignoring the
terms with $W_{2}$ and $W_{3}$) one regains after performing the thermal
averaging, the PGH kernel%
\begin{equation}
P_{0}\left( E^{\prime }|E\right) =\left( \frac{\beta }{4\pi DE_{1}}\right)
^{1/2}\exp \left( -\frac{\beta \left( E^{\prime }-E+DE_{1}\right) ^{2}}{%
4DE_{1}}\right) .  \label{4.7}
\end{equation}%
One can also explicitly perform the averaging over all modes with the full
expression (including the terms with $W_{2}$ and $W_{3}$), since it involves
only Gaussian integrations however we have not found a way to express the
final result in the continuum limit. Instead we will consider the lowest
order perturbation theory limit.

For this purpose we note that the conditional probability, to lowest order
has to obey three conditions. The first is normalization, that is%
\begin{equation}
\int_{-\infty }^{\infty }dE^{\prime }P\left( E^{\prime }|E\right) =1.
\label{4.8}
\end{equation}%
The second is that the average change in energy must be given by the average energy
change found in the previous section, that is%
\begin{equation}
\left\langle \Delta E\right\rangle =\int_{-\infty }^{\infty }dE^{\prime
}P\left( E^{\prime }|E\right) \left( E-E^{\prime }\right) =DE_{1}-DE_{2}.
\label{4.9}
\end{equation}%
Thirdly, the conditional probability must obey detailed balance, that is%
\begin{equation}
P\left( E^{\prime }|E\right) \exp \left( -\beta E\right) =P\left(
E|E^{\prime }\right) \exp \left( -\beta E^{\prime }\right) .  \label{4.10}
\end{equation}

To simplify, we use henceforth the following reduced notations:%
\begin{equation}
\varepsilon =\beta E,\text{ \ }\delta _{1}=\beta DE_{1},\text{ \ }\delta
_{2}=\beta DE_{2},\text{ }y=\varepsilon ^{\prime }-\varepsilon .\text{\ }
\label{4.11}
\end{equation}%
The leading order correction to the conditional probability, which also
obeys detailed balance must be of the order $y^{2}$. We then write the
conditional probability as:%
\begin{equation}
P\left( \varepsilon ^{\prime }|\varepsilon \right) =a\left( 1-by^{2}\right)
P_{0}\left( \varepsilon ^{\prime }|\varepsilon \right) .  \label{4.12}
\end{equation}%
The two coefficients are determined by the normalization condition and the
known thermally averaged energy loss:%
\begin{eqnarray}
1 &=&\int_{-\infty }^{\infty }dya\left( 1-by^{2}\right) \left( \frac{1}{4\pi
\delta _{1}}\right) ^{1/2}\exp \left( -\frac{\left( y+\delta _{1}\right) ^{2}%
}{4\delta _{1}}\right)   \label{4.13} \\
\delta _{1}-\delta _{2} &=&-\int_{-\infty }^{\infty }dyya\left(
1-by^{2}\right) \left( \frac{1}{4\pi \delta _{1}}\right) ^{1/2}\exp \left( -%
\frac{\left( y+\delta _{1}\right) ^{2}}{4\delta _{1}}\right) .  \label{4.14}
\end{eqnarray}%
One then finds that:%
\begin{equation}
P\left( \varepsilon ^{\prime }|\varepsilon \right) =\left( 1+\frac{\delta
_{2}}{4\delta _{1}}\left( 2+\delta _{1}-\frac{y^{2}}{\delta _{1}}\right)
\right) P_{0}\left( \varepsilon ^{\prime }|\varepsilon \right) .
\label{4.15}
\end{equation}

\subsection{\protect\bigskip The depopulation factor}

The depopulation factor given in the turnover theory is
\begin{equation}
\Upsilon =\exp \left( \frac{1}{2\pi }\int_{-\infty }^{\infty }d\tau \frac{%
\ln \left[ 1-\tilde{P}\left( \tau -\frac{i}{2}\right) \right] }{\tau ^{2}+%
\frac{1}{4}}\right) .  \label{4.16}
\end{equation}%
where $\tilde{P}\left( \tau -\frac{i}{2}\right) $ is the Fourier transform
of the conditional probability kernel. One finds:%
\begin{eqnarray}
\tilde{P}\left( \tau -\frac{i}{2}\right)  &=&\int_{-\infty }^{\infty }dy\exp
\left( i(\tau -\frac{i}{2})y\right) P\left( y\right)   \notag \\
&=&\left( 1+\delta _{2}\left( \tau ^{2}+\frac{1}{4}\right) \right) \exp
\left( -\delta _{1}\left( \tau ^{2}+\frac{1}{4}\right) \right) .
\label{4.17}
\end{eqnarray}%
Assuming that the correction is small, the depopulation factor is then
expanded to%
\begin{equation}
\Upsilon \simeq \Upsilon _{0}\exp \left( -\frac{2\delta _{2}}{\delta _{1}}%
\left( 1-\rm{erf}\left( \frac{\sqrt{\delta _{1}}}{2}\right) \right)
\right)   \label{4.18}
\end{equation}%
where $\Upsilon _{0}$ is the depopulation factor using the zero-th order
kernel in Eq. \ref{4.16}. One notes that the thermal correction $\delta
_{2}$ reduces the depopulation factor and thus the rate. The reduced energy
loss of the particle, effectively makes it more difficult to energize the
particle and the rate is reduced. As we shall see considering a specific
example below, the ratio $\frac{\delta _{2}}{\delta _{1}}$ is typically of
the order of $1/(\beta V^{\ddag })$ so that it is small and does not affect
the barrier crossing rate appreciably.

\renewcommand{\theequation}{5.\arabic{equation}} \setcounter{section}{4} %
\setcounter{equation}{0}

\section{An example - Ohmic friction and escape from a cubic potential well}

For Ohmic friction,
the
memory kernel in the normal mode representation as taken from Ref. \cite{rips90} is:%
\begin{equation}
K\left( t\right) =\frac{u_{00}^{2}}{2\lambda ^{\ddag ^{2}}}\left( \exp
\left( -\lambda ^{\ddag }t\right) -\frac{\lambda ^{\ddag }}{\lambda _{1}}%
\exp \left( -\lambda _{1}t\right) +1+\frac{\lambda ^{\ddag }}{\lambda _{1}}%
\right) -\frac{1}{\omega ^{\ddag ^{2}}}  \label{5.2}
\end{equation}%
with
\begin{equation}
\lambda _{1}=\left( \frac{\gamma ^{2}}{4}+\omega ^{\ddag ^{2}}\right) ^{1/2}+%
\frac{\gamma }{2}.  \label{5.3}
\end{equation}%
For a cubic potential%
\begin{equation}
V\left( q\right) =-\frac{M\omega ^{\ddag ^{2}}}{2}q^{2}\left( 1+\frac{q}{%
q_{0}}\right)   \label{5.4}
\end{equation}%
one finds \cite{rips90} that the solution for the zero-th order equation of motion for the
unstable mode (Eq. \ref{2.16a}) is:%
\begin{equation}
\rho _{0,t}=-\frac{\lambda ^{\ddag ^{2}}\sqrt{M}q_{0}}{u_{00}^{3}\omega
^{\ddag ^{2}}\cosh ^{2}\left( \frac{\lambda ^{\ddag }}{2}t\right) }
\label{5.5}
\end{equation}%
so that%
\begin{eqnarray}
DE_{2} &=&\frac{36}{\beta }\int_{-\infty }^{\infty }d\bar{x}\int_{0}^{\infty
}d\Delta x\left( 1-\exp \left( -2\Delta x\frac{\gamma }{\lambda ^{\ddag }}%
\right) \right) \exp \left( -2\Delta x\right)   \notag \\
&&\cdot \frac{\sinh \left( \bar{x}+\frac{\Delta x}{2}\right) }{\cosh
^{5}\left( \bar{x}+\frac{\Delta x}{2}\right) }\frac{\sinh \left( \bar{x}-%
\frac{\Delta x}{2}\right) }{\cosh ^{5}\left( \bar{x}-\frac{\Delta x}{2}%
\right) }\left[ M\left( \bar{x}+\frac{\Delta x}{2}\right) -M\left( \bar{x}-%
\frac{\Delta x}{2}\right) \right]   \label{5.6}
\end{eqnarray}%
where we used the notation%
\begin{equation}
\int_{\frac{\lambda ^{\ddag }}{2}t^{\prime \prime }}^{\frac{\lambda ^{\ddag }%
}{2}t}dx\frac{\cosh ^{6}x}{\sinh ^{2}x}=M\left( \frac{\lambda ^{\ddag }}{2}%
t\right) -M\left( \frac{\lambda ^{\ddag }}{2}t^{\prime \prime }\right) .
\label{5.7}
\end{equation}
In the limit of weak friction $\frac{\gamma }{\lambda ^{\ddag }}\ll 1$ this
reduces to
\begin{equation}
DE_{2}\left( \gamma \rightarrow 0\right) \simeq \frac{5\gamma }{\beta \omega
^{\ddag }}.  \label{5.8}
\end{equation}%
In this limit, the temperature independent part of the average energy loss is \cite{rips90}%
\begin{equation}
DE_{1}\left( \gamma \rightarrow 0\right) \simeq \frac{36}{5}\frac{\gamma
V^{\neq }}{\omega ^{\ddag }}  \label{5.9}
\end{equation}%
so that the ratio of the two terms is%
\begin{equation}
\frac{DE_{2}\left( \gamma \rightarrow 0\right) }{DE_{1}\left( \gamma
\rightarrow 0\right) }\simeq \frac{25}{36\beta V^{\neq }}  \label{5.10}
\end{equation}%
which is a small number. As noted above, the temperature induced correction
may be ignored in this limit.

It is also of interest to estimate the temperature dependent energy gain in
the opposite limit that is when $\frac{\gamma }{\lambda ^{\ddag }}\simeq
\frac{\gamma ^{2}}{\omega ^{\ddag ^{2}}}\gg 1$. One readily finds that
\begin{equation}
DE_{2}\left( \gamma \rightarrow \infty \right) \simeq \frac{5}{\beta }.
\label{5.11}
\end{equation}%
In this same limit the temperature independent energy loss is \cite{rips90}%
\begin{equation}
DE_{1}\left( \gamma \rightarrow \infty \right) =\frac{27}{35}V^{\neq }
\label{5.12}
\end{equation}%
so that:%
\begin{equation}
\frac{DE_{2}\left( \gamma \rightarrow \infty \right) }{DE_{1}\left( \gamma
\rightarrow \infty \right) }=\frac{175}{27\beta V^{\neq }}  \label{5.13}
\end{equation}%
and this is not necessarily small. Strictly speaking, in this
limit the perturbation expansion is invalid, however the result does point
out that when the friction is not too weak one may find discrepancies
between the "standard"\ PGH theory energy loss and the true energy loss, which originate from
the fact that the thermal bath can also contribute to decreasing the energy
loss of the system.

\begin{figure}[htb]
  \centering  	
    \includegraphics[width=0.6\textwidth]{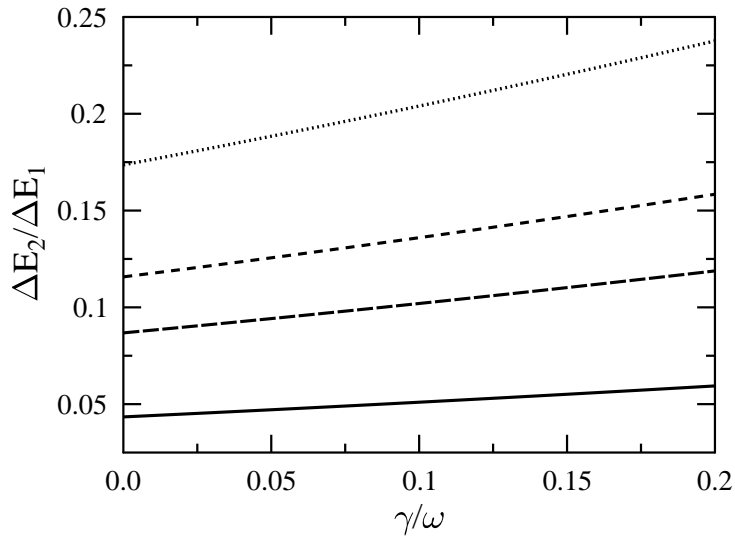}
     \caption{Frictional dependence of the ratio of the thermally induced energy gain (\ref{5.6}) and the energy loss (\ref{2.29}) for increasing barrier height (top to bottom): $\beta V^{\neq}$= 4 (dotted), 6 (short-dashed), 8 (dashed), 16 (solid).}\label{fig:loss}
\end{figure}
\begin{figure}[htb]
  \centering  	
    \includegraphics[width=0.6\textwidth]{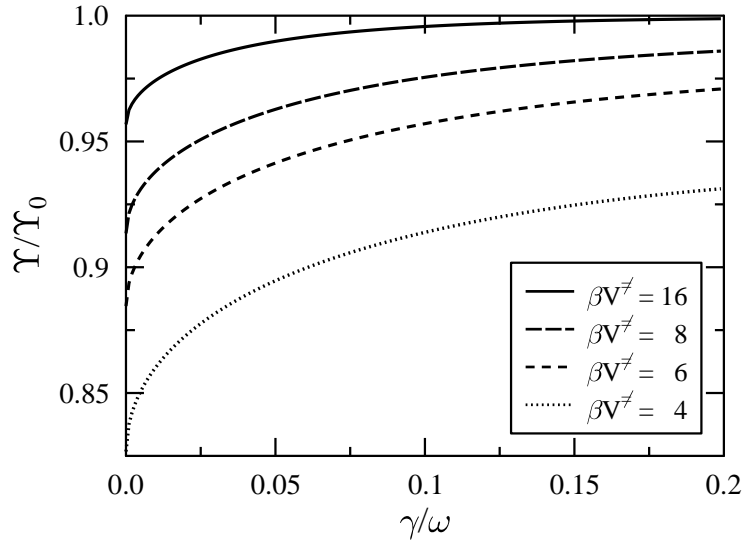}
     \caption{Ratio of the depopulation factors $\Upsilon/\Upsilon_0$ according to (\ref{4.16}) with (\ref{4.15}) and without (\ref{4.7}) the thermal correction to the energy transfer kernel.}\label{fig:rateratio}
\end{figure}

The above results are illustrated in Figs.~\ref{fig:loss}, \ref{fig:rateratio} where we show the thermally induced energy gain relative to the energy loss and the reduction of the rate due to a reduction of the depopulation factor (\ref{4.16}) compared to the PGH result. According to the above discussion, in the domain of low friction the ratio $DE_2/DE_1$ is mainly determined by $\beta V^{\neq}$ with the tendency to be more sensitive to dissipation for lower (reduced) barrier heights. The impact of the energy gain on the depopulation factor decreases with growing friction and increasing barrier height as indicated in (\ref{4.18}):  for fixed barrier height $\delta_2/\delta_1$ is almost constant while $\delta_1\sim \gamma V^{\neq}$ grows so that $\Upsilon/\Upsilon_0\to 1$.
  In typical experimental situations with $\beta V^{\neq}$ of the order of 10, small deviations from the PGH prediction may be found in the weak damping regime $\gamma/\omega < 0.1$.

\section{Discussion}

The solution of the Kramers turnover problem in the 1980s has been a cornerstone result of thermal rate theory \cite{htb,pollak05}. The rate expression obtained within PGH theory has been of practical use in a variety of fields such as chemical reactions \cite{muller08,muller12} and Josephson junction physics \cite{jj1,jj2}. However, it is incomplete as we have shown in this work since it does not include the fact that the temperature of the surrounding bath degrees of freedom will typically reduce the average energy lost by the system, due to the thermal fluctuations. Within a consistent perturbative treatment, this "heating" contribution should be taken into account on the same footing as the dissipative part. One then finds that the thermal fluctuations indeed lead to a reduction of the energy loss that a particle experiences when traversing the well of a barrier potential. In contrast, the variance is affected only in higher order of perturbation theory. Consequently, the depopulation factor of the rate is reduced compared to the PGH prediction.

This result cannot be obtained within the Mel'nikov and Meshkov formulation where the heating contribution diverges. While  the corresponding depopulation factor has often been used in practical calculations due to its simpler structure, our analysis demonstrates
 the inconsistency of the approach. Discrepancies between the corrected rate expression of the PGH theory and the original one are expected to be detectable in the low friction regime and in systems with moderate (reduced) barrier heights.

 In this paper we have limited ourselves to the lowest order consistent contribution to the rate expression. Within the PGH framework, it is possible to go to higher order and thus obtain a further temperature dependent consistent correction also to the variance of the energy transfer. One may use the same perturbation theory to also obtain improved dynamical estimates to the rate, using the formalism of Ref. \cite{pollak93}.

\section*{Acknowledgements}

This work was supported by a grant of the German Israel Foundation for Basic
Research. EP thanks the Alexander von Humboldt Foundation for a continuation of his senior fellowship, during which part of this work was initiated.

\renewcommand{\theequation}{A.\arabic{equation}} \setcounter{equation}{0}

\appendix*
\section{Unstable mode motion in first order}

In this appendix we note that by employing energy conservation of the
composite system and bath one obtains a first order in time equation of
motion for the first order correction to the unstable mode motion. Using the
representation of the Hamiltonian in terms of the normal modes (Eq. \ref{2.7}%
) one readily finds that to first order%
\begin{equation}
p_{\rho 1,t}p_{\rho 0,t}=\lambda ^{\ddag 2}\rho _{0,t}\rho _{1,t}-\frac{%
u_{00}\rho _{1,t}+\Phi \left( t\right) }{\sqrt{M}}V_{1}^{\prime }\left(
\frac{u_{00}\rho _{0,t}}{\sqrt{M}}\right) -\sum_{j=1}^{N}\left[
p_{y_{j1,t}}p_{y_{j0,t}}+\lambda _{j}^{2}y_{j1,t}y_{j0,t}\right] .
\label{A.1}
\end{equation}%
Noting that
\begin{equation}
\dot{\rho}_{1,t}=p_{\rho 1,t}  \label{A.2}
\end{equation}
we obtain a first order in time equation of motion%
\begin{equation}
\dot{\rho}_{1,t}=\frac{d}{dt}\left[ \ln \dot{\rho}_{0,t}\right] \rho _{1,t}-%
\frac{\Phi \left( t\right) }{\dot{\rho}_{0,t}\sqrt{M}}V_{1}^{\prime }\left(
\frac{u_{00}\rho _{0,t}}{\sqrt{M}}\right) -\frac{1}{\dot{\rho}_{0,t}}%
\sum_{j=1}^{N}\left[ p_{y_{j1,t}}p_{y_{j0,t}}+\lambda
_{j}^{2}y_{j1,t}y_{j0,t}\right] .  \label{A.3}
\end{equation}%
The solution for the first order correction to the motion of the stable
modes has been given in Eq. \ \ref{2.24}. Noting that at the initial time $%
t_{0}$ the potential term $V_{1}\left( \frac{u_{00}\rho _{0,t}}{\sqrt{M}}%
\right) $ vanishes, one finds that the solution for the first order
term for the unstable mode motion is:%
\begin{equation}
\rho _{1,t}=-\dot{\rho}_{0,t}\int_{t_{0}}^{t}dt^{\prime }\frac{1}{\dot{\rho}%
_{0,t^{\prime }}^{2}}\int_{t_{0}}^{t^{\prime }}dt^{\prime \prime }\frac{\Phi
\left( t^{\prime \prime }\right) }{\sqrt{M}}\frac{d}{dt^{\prime \prime }}%
V_{1}^{\prime }\left( \frac{u_{00}\rho _{0}\left( t^{\prime \prime }\right)
}{\sqrt{M}}\right)   \label{A.4}
\end{equation}%
and it is linear in the noise $\Phi $. It is an instructive exercise to
show by direct differentiation with respect to the time that this solution
is also a solution of the second order equation of motion, as given in Eq. %
\ref{2.17}.

One may follow this procedure in principle to also obtain the higher order
solutions. Specifically, given $\rho _{1,t}$ one obtains an explicit result
for the second order contribution to the j-th stable mode $y_{j2,t}$. Then
one uses again the conservation of energy to obtain a first order in time
equation for $\rho _{2,t}$ which is readily solved, and so on.

\end{document}